\documentclass[12pt]{article}

\usepackage{amsmath}
\usepackage{amssymb}
\usepackage{latexsym}
\usepackage{mathrsfs}
\usepackage{amsfonts}

\newtheorem{thm}{Theorem}
\newtheorem{prop}{Proposition}

\newtheorem{defn}{Definition}

\newtheorem{lemma}{Lemma}
\def\RR{{\bf R}}

\def\<{\langle}
\def\>{\rangle}
\def\pf{\noindent{\bf Proof.} }

\def\qed{{\hfill $\Box$\medskip}}

\def\to{\rightarrow}
\def\QQ{\mathcal{Q}}

\def\PP{\mathbb{P}}
\def\EE{\mathbb{E}}
\def\NN{\mathbb{N}}

\def\cvar{\mathrm{CVaR}}
\def\var{\mathrm{VaR}}
\def\LL{\mathscr{L}}

\def\RR{\mathbb{R}}
\def\RRR{\mathcal{R}}
\def\cl{\mathrm{cl}}

\def\maxvar{\mathrm{MAXVAR}}
\def\minvar{\mathrm{MINVAR}}

\def\var{\mathrm{VaR}}
\allowdisplaybreaks[4]

\input{epsf.sty}
\usepackage{epsfig}

\begin{document}
\title{\bf On coherency and other properties of MAXVAR
\footnote{ This paper is dedicated to Michel Th\'era in celebration of his 70th birthday.}}
\author{Jie Sun\footnote{School of Mathematical Sciences, Chongqing Normal University, PRC and School of Science and CIC, Curtin University, Australia. Email: jie.sun@curtin.edu.au.}~~and~Qiang Yao\footnote{School of Statistics, East China Normal University, PRC and NYU---ECNU Institute of Mathematical Sciences at NYU Shanghai, China.
                      E-mail: qyao@sfs.ecnu.edu.cn.}
        }
%\date{}

\maketitle{}

\begin{abstract}
This paper is concerned with the MAXVAR risk measure on $\LL^2$ space. We present an elementary and direct proof of its coherency and averseness. Based on the observation that the MAXVAR measure is a continuous convex combination of the CVaR measure, we provide an explicit formula for the risk envelope of MAXVAR.

 \end{abstract} \noindent{\bf 2010 MR subject classification:}~49N15, 91G40


\section{Introduction}
In Cherny and Madan \cite{Cherny-Madan2009} and Cherny and Orlov \cite{Cherny-Orlov2009}, a new kind of risk measure --``MAXVAR'' -- is proposed, which is useful in the analysis of large portfolios. Given a probability space $(\Omega,\Sigma,\PP_0)$  and a random variable $X\in\LL^2(\Omega, \Sigma, \PP_0)$, where $ \LL^2(\Omega, \Sigma, \PP_0)$ is the square integrable Lebesgue space ($\LL^2$ for short), the  MAXVAR is defined as

$$\maxvar_n(X):=\EE(\max\{X_1,\cdots,X_n\}),$$
where $X_1,\cdots,X_n$ are i.i.d. copies of $X$. We call $\maxvar_n(\cdot)$ the ``MAXVAR risk measure''.

Note that $\maxvar_n(\cdot)$ is always finite  on $\LL^2$ since $|\maxvar_n(X)|\leq n\EE(|X|)<+\infty$ for any $X\in\LL^2$.

In  \cite{Cherny-Madan2009,Cherny-Orlov2009},  the name of ``MINVAR risk measure'' was used. Since we treat risk measures as a nondecreasing function, we use ``MAXVAR risk measure'' instead. Obviously, we have $\maxvar_n(X)=-\minvar_n(-X)$. Different from papers \cite{Cherny-Madan2009,Cherny-Orlov2009}, which considered coherency of MINVAR in $\LL^\infty$ space, this paper  deals with the $\LL^2$ space. Our proof of the coherency of MAXVAR risk measure is direct and independent of  \cite{Cherny-Madan2009,Cherny-Orlov2009}. Moreover, we show  risk averseness of MAXVAR and give an explicit formula for its risk envelope.\\

In Section 2, we present a simple proof for the coherency  of MAXVAR. We show its aversity in Section 3. Section 4 is devoted to the discussion of a continuous representation of MAXVAR  and Section 5 provides an explicit formula for its risk envelope.

\section{Coherency of MAXVAR}

In this section, we show that  MAXVAR  is a coherent risk measure in basic sense of Rockafellar.
\begin{defn}\label{d:coherent} (Rockafellar \cite{Rockafellar2007}) A functional $\RRR:\LL^2\to(-\infty,+\infty]$ is a coherent risk measure in basic sense if it satisfies
\begin{description}
  \item[(A1)] $\RRR(C)=C$ for all constant $C$;
  \item[(A2)] (``convexity'')~$\RRR(\lambda X+(1-\lambda)Y)\leq\lambda\cdot\RRR (X)+(1-\lambda)\cdot\RRR(Y)$ for any $X,Y\in\LL^2$ and any fixed $0\leq\lambda\leq1$;
  \item[(A3)] (``monotonicity'')~$\RRR(X)\leq\RRR(Y)$ for any $X,Y\in\LL^2$ satisfying $X\leq Y$;
  \item[(A4)] (``closedness'')~If $\|X^k-X\|_2\rightarrow0$ and $\RRR(X^k)\leq0$ for all $k\in\NN$, then $\RRR(X)\leq0$;
  \item[(A5)] (``positive homogeneity'')~$\RRR(\lambda X)=\lambda\RRR(X) $ for any $\lambda>0$ and $X\in\LL^2$.
\end{description}
\end{defn}

\bigskip

\begin{thm}\label{t:coherent}
$\maxvar_n(\cdot)$ is a coherent risk measure in basic sense.
\end{thm}

\pf {\bf(A1)} is obvious by definition. {\bf (A5)} is also easy to check since if $X_1,\cdots,X_n$ are i.i.d. copies of $X$ and $\lambda>0$, then $\lambda X_1,\cdots,\lambda X_n$ are i.i.d. copies of $\lambda X$.\\

\noindent{\it Proof of {\bf(A2)}.} We only need to show the following subadditive property of $\maxvar$
\begin{equation}\label{subadd}\maxvar_n(X+Y)\le\maxvar_n(X)+\maxvar_n(Y)\quad \forall X,Y.\end{equation}
Then (\ref{subadd}) and {\bf(A5)} imply {\bf(A2)}.
For any $X,Y\in\LL^2$, take $(X_1,Y_1),\cdots,(X_n,Y_n)$ as i.i.d. copies of the two dimensional random vector $(X,Y)$. That is, the random vectors $(X_1,Y_1),\cdots,(X_n,Y_n)$ are independent and have the same joint distribution as the random vector $(X,Y)$. Then $X_1,\cdots,X_n$ are i.i.d. copies of $X$ and $Y_1,\cdots,Y_n$ are i.i.d. copies of $Y$. We next show that  $X_1+Y_1,\cdots,X_n+Y_n$ are i.i.d. copies of $X+Y$.

Since $(X_i,Y_i)$ has the same joint distribution as $(X,Y)$, $i=1,...,n$, it follows that $X_i+Y_i$ has the same distribution as $X+Y$. In order to prove that $X_1+Y_1,\cdots,X_n+Y_n$ are independent, we only need to prove that for any $t_1,\cdots,t_n\in\RR$,
\begin{align}\label{e:1}
&\PP_0\left(X_1+Y_1\leq t_1,\cdots,X_n+Y_n\leq t_n\right)\nonumber\\=&\PP_0\left(X_1+Y_1\leq t_1\right)\cdot\cdots\cdot\PP_0\left(X_n+Y_n\leq t_n\right).
\end{align}
In fact, since the random vectors $(X_1,Y_1),\cdots,(X_n,Y_n)$ are independent, we have
\begin{align}\label{e:2}
&\PP_0((X_1,Y_1)\in B_1,\cdots,(X_n,Y_n)\in B_n)\nonumber\\=&\PP_0((X_1,Y_1)\in B_1)\cdot\cdots\cdot\PP_0((X_n,Y_n)\in B_n)
\end{align}
for any Borel sets $B_1,\cdots,B_n\subseteq\RR^2$. In particular, if we take $$B_i=\left\{(x,y)\in\RR^2:~{x+y}\leq t_i\right\}$$ for any $1\leq i\leq n$ in (\ref{e:2}), we can get (\ref{e:1}). Therefore, $X_1+Y_1,\cdots,X_n+Y_n$ are independent. Moreover, they are i.i.d. copies of $X+Y$.

Since the definition of MAXVAR does not depend on the choice of the i.i.d. copies, we have
\begin{align*}
\maxvar_n(X)&=\EE(\max\{X_1,\cdots,X_n\}),\\
\maxvar_n(Y)&=\EE(\max\{Y_1,\cdots,Y_n\}),\\
\maxvar_n(X+Y)&=\EE(\max\{X_1+Y_1,\cdots,X_n+Y_n\}).
\end{align*}
Furthermore, since
$$\max\{ X_1+Y_1,\cdots, X_n+Y_n\}\leq\max\{X_1,\cdots,X_n\}+\max\{Y_1,\cdots,Y_n\},$$
we get
\begin{align*}
\maxvar_n( X+Y)&=\EE(\max\{X_1+Y_1,\cdots,X_n+Y_n\})\\
&\leq\EE(\max\{X_1,\cdots,X_n\})+\EE(\max\{Y_1,\cdots,Y_n\})\\
&=\maxvar_n(X)+\maxvar_n(Y).
\end{align*}

\noindent {\it Proof of {\bf(A3)}}. For any $X,Y\in\LL^2$ satisfying $X\leq Y$, suppose $X_1,\cdots,X_n$ are i.i.d. copies of $X$ and $Y_1,\cdots,Y_n$ are i.i.d. copies of $Y$. We can see that $\PP_0(X\leq t)\geq\PP_0(Y\leq t)$ for any $t\in\RR$ since $X\leq Y$. Then we have
\begin{align}
&\maxvar_n(X)\nonumber\\
&=\int_{-\infty}^0\left[\PP_0(\max\{X_1,\cdots,X_n\}>t)-1\right]dt+\int_0^{+\infty}\PP_0(\max\{X_1,\cdots,X_n\}>t)dt\nonumber\\
&=-\int_{-\infty}^0(\PP_0(X\leq t))^ndt+\int_0^{+\infty}\left[1-(\PP_0(X\leq t))^n\right]dt\nonumber\\
&\leq-\int_{-\infty}^0(\PP_0(Y\leq t))^ndt+\int_0^{+\infty}\left[1-(\PP_0(Y\leq t))^n\right]dt\nonumber\\
&=\int_{-\infty}^0\left[\PP_0(\max\{Y_1,\cdots,Y_n\}>t)-1\right]dt+\int_0^{+\infty}\PP_0(\max\{Y_1,\cdots,Y_n\}>t)dt\nonumber\\
&=\maxvar_n(Y).\nonumber
\end{align}
The detail of the first equality is as follows. Denote by $$F(t)=\PP_0(\max\{X_1,\cdots,X_n\}\leq t)$$ the cumulative distribution function of $\max\{X_1,\cdots,X_n\}$. Then
\begin{eqnarray}
\EE(\max\{X_1,\cdots,X_n\})&=&\int_{-\infty}^{+\infty}xdF(x)\nonumber\\
&=&-\int_{-\infty}^0\left[\int_x^0dt\right]dF(x)+\int_0^{+\infty}\left[\int_0^xdt\right]dF(x)\nonumber\\
\hbox{(by Fubini's Theorem)}&=&-\int_{-\infty}^0\left[\int_{-\infty}^tdF(x)\right]dt+\int_0^{+\infty}\left[\int_t^{+\infty}dF(x)\right]dt\nonumber\\
&=&-\int_{-\infty}^0F(t)dt+\int_0^{+\infty}[1-F(t)]dt.\label{e:2.1}
\end{eqnarray}
And the second equality comes from the fact that $F(t)=(\PP_0(X\leq t))^n$.\\

\noindent {\it Proof of {\bf(A4)}.}  Suppose $X^k~(k=1,2,\cdots),X\in\LL^2$ and $\|X^k-X\|_2\rightarrow0$ as $k$ tends to infinity. Then $X^k\rightarrow X$ in distribution. Denote by $F_k(t)$ the distribution function of $X^k~(k=1,2,\cdots)$ and by $F(t)$ the distribution of $X$. Then $\lim\limits_{k\rightarrow\infty}F_k(t)=F(t)$ for all continuous points of $F(\cdot)$. It implies that
$\lim\limits_{k\rightarrow\infty}[F_k(t)]^n=[F(t)]^n$ for all continuous points of $[F(\cdot)]^n$. Note that $[F_k(t)]^n$ is the distribution function of $\max\{X^k_1,\cdots,X^k_n\}$ and $[F(t)]^n$ is the distribution function of $\max\{X_1,\cdots,X_n\}$, where $X^k_1,\cdots,X^k_n$ are i.i.d. copies of $X^k~(k=1,2,\cdots)$ and $X_1,\cdots,X_n$ are i.i.d. copies of $X$. Therefore, we have $\max\{X^k_1,\cdots,X^k_n\}\rightarrow\max\{X_1,\cdots,X_n\}$ in distribution, and
\begin{eqnarray*}\maxvar_n(X^k)&=&\EE(\max\{X^k_1,\cdots,X^k_n\})\\&\rightarrow&\EE(\max\{X_1,\cdots,X_n\})=\maxvar_n(X)
\end{eqnarray*} as $k$ tends to infinity. Thus, if $\maxvar_n(X^k)\leq0$ for all $k=1,2,\cdots$ then $\maxvar_n(X)\leq0$. The proof of the theorem is completed.\qed

\section{Risk-averseness of MAXVAR}

Suppose $\RRR$ is a functional from $\LL^2$ to $(-\infty,+\infty]$. Recall that an \emph{averse} risk measure is defined by axioms (A1), (A2), (A4), (A5) and
\begin{description}
  \item[(A6)] $\RRR(X)>\EE(X)$ for all non-constant $X$.
\end{description}

We then have the next theorem.
\begin{thm}\label{t:averse}
If $n\geq2$, then $\maxvar_n(\cdot)$ is averse.\\
\end{thm}

F\"ollmer and Schied \cite{Follmer-Schied2002}  proved that if $\RRR$ is a coherent, law-invariant risk measure in $\LL^\infty$~(not $\LL^2$) other than $\EE(\cdot)$, then $\RRR$ is averse, where ``law-invariant'' stands for that $\RRR(X)=\RRR(Y)$ whenever $X$ and $Y$ have the same distribution under $\PP_0$. Since we are now considering the $\LL^2$ case, we cannot use the result in F\"ollmer and Schied \cite{Follmer-Schied2002} directly. We next give a separate proof.\\

\noindent\textbf{\it Proof of Theorem \ref{t:averse}}~On one hand, for any $X\in\LL^2$, let $X_1,\cdots,X_n$ be i.i.d. copies of $X$. Then we have
$$\maxvar_n(X)=\EE(\max\{X_1,\cdots,X_n\})\geq\EE(X_1)=\EE(X).$$ On the other hand, if $\maxvar_n(X)=\EE(X)=\EE(X_1)~(n\geq2)$, then \newline $\max\{X_1,\cdots,X_n\}=X_1$ almost surely. Similarly, $\max\{X_1,\cdots,X_n\}=X_2$ almost surely. Therefore, $X_1=X_2$ almost surely. Since $X_1$ and $X_2$ are independent, we must have $X_1$ equals to a constant almost surely, which is equivalent to say $X$ equals to a constant almost surely. Therefore, $\maxvar_n(X)>\EE(X)$ for nonconstant $X$, which implies that $\maxvar_n(\cdot)$ is averse when $n\geq2$.\\

\noindent\textbf{Remark.}~In fact, Theorems 1 and 2 can be obtained as corollaries of Theorem 3 in the next section. See the remark after the proof of Theorem 3 for details. However, we think it is of interest to provide an elementary proof only based on  definition of MAXVAR.

\section{MAXVAR as a continuous convex combination of CVaR}
An important coherent risk measure in basic sense is the conditional value at risk (CVaR) popularized by Rockafellar and Uryasev \cite{Rockafellar-Uryasev2000}. Among several equivalent definitions of CVaR, the most familiar one is probably the following.
\begin{equation}\label{32}{\rm CVaR}_\alpha(X)=\min_{\beta\in\RR}\left\{  \beta+{1\over 1-\alpha}\EE(X-\beta)_+\right\},\end{equation} where $(t)_+=\max(t,0)$ and $\alpha\in[0,1)$. The minimum is attained at $\beta^*=\var_\alpha(X)$, and the VaR ( ``Value-at-Risk")  is defined as
\begin{equation}\label{33}\var_\alpha(X):=\inf\left\{\nu\in\RR:\PP_0(X>\nu)<1-\alpha\right\}.
\end{equation}

In this section, we show that  $\maxvar_n(\cdot)$ is certain ``continuous convex combination'' of the  CVaR measure in the sense that
$$\maxvar_n(\cdot)=\int_0^1\cvar_\alpha(\cdot)w_n(\alpha)d\alpha,$$ where $w_n(\alpha)~(\alpha\in[0,1])$ is the ``weight function'' which satisfies $w_n(\alpha)\geq0$ on $[0,1]$ and $\int_0^1w_n(\alpha)d\alpha=1$. Specifically, we have the next theorem.
\begin{thm}\label{t:mixedCVaR}
For any $X\in\LL^2$, we have $$\maxvar_n(X)=\int_0^1\cvar_\alpha(X)w_n(\alpha)d\alpha,$$ where $$w_n(\alpha):=n(n-1)(1-\alpha)\alpha^{n-2},~~~~~~\alpha\in[0,1]$$ is the weight function.
\end{thm}

\noindent\textbf{Remark.}~It can be easily checked that $w_n(\alpha)\geq0$ on $[0,1]$, and
$$\int_0^1w_n(\alpha)d\alpha=n(n-1)\int_0^1(\alpha^{n-2}-\alpha^{n-1})d\alpha=n(n-1)\left[\frac{1}{n-1}-\frac{1}{n}\right]=1.$$ Therefore, $w_n(\alpha)$ is indeed a weight function.\\

Theorem \ref{t:mixedCVaR} was mentioned in Cherny and Orlov \cite{Cherny-Orlov2009} without details. We now give a detailed  proof by using the so called ``Choquet integral''. First, we need a lemma. For any $\alpha\in[0,1)$, define $f_\alpha(\cdot):~\Sigma\rightarrow[0,1]$ in the following way,
\begin{align*}
f_\alpha(A):&=\left\{\begin{array}{ll}\frac{1}{1-\alpha}\PP_0(A)~~~~~~~\text{if}~\PP_0(A)\leq1-\alpha,\\~~~~~~1~~~~~~~~~~~~~\text{otherwise}. \end{array}\right.\\
&=g_\alpha[\PP_0(A)],
\end{align*}
where
\begin{equation}\label{e:3.1}
g_\alpha(x):=\left\{\begin{array}{ll}\frac{1}{1-\alpha}x~~~~~~~\text{if}~x\in[0,1-\alpha),\\~~~1~~~~~~~~~~\text{if}~x\in[1-\alpha,1]. \end{array}\right.
\end{equation}

We then have the following lemma, which implies that the CVaR measure can be written as the ``Choquet integral'' with respect to $f_\alpha(\cdot)$.
\begin{lemma}\label{l:Choquet}
For any $X\in\LL^2$ and $\alpha\in[0,1)$, we have
$$\cvar_\alpha(X)=\int_{-\infty}^0[f_\alpha(X>t)-1]dt+\int_0^{+\infty}f_\alpha(X>t)dt.$$
\end{lemma}
\pf If $\var_\alpha(X)\leq0$, then
\begin{align*}
&\int_{-\infty}^0[f_\alpha(X>t)-1]dt+\int_0^{+\infty}f_\alpha(X>t)dt\\
=&\int_{\var_\alpha(X)}^0\left[\frac{1}{1-\alpha}\PP_0(X>t)-1\right]dt+\int_0^{+\infty}\frac{1}{1-\alpha}\PP_0(X>t)dt\\
=&\var_\alpha(X)+\frac{1}{1-\alpha}\cdot\int_{\var_\alpha(X)}^{+\infty}\PP_0(X>t)dt\\
=&\var_\alpha(X)+\frac{1}{1-\alpha}\cdot\EE[(X-\var_\alpha(X))_+]=\cvar_\alpha(X).
\end{align*}
The last step above is due to (\ref{32}) and (\ref{33}).

If $\var_\alpha(X)>0$, then
\begin{align*}
&\int_{-\infty}^0[f_\alpha(X>t)-1]dt+\int_0^{+\infty}f_\alpha(X>t)dt\\
=&\int_0^{\var_\alpha(X)}dt+\int_{\var_\alpha(X)}^{+\infty}\frac{1}{1-\alpha}\PP_0(X>t)dt\\
=&\var_\alpha(X)+\frac{1}{1-\alpha}\cdot\int_{\var_\alpha(X)}^{+\infty}\PP_0(X>t)dt\\
=&\var_\alpha(X)+\frac{1}{1-\alpha}\cdot\EE[(X-\var_\alpha(X))_+]=\cvar_\alpha(X),
\end{align*}
which completes the proof.\qed

\noindent\textbf{\it Proof of Theorem \ref{t:mixedCVaR}}\quad Define $$h(x):=1-(1-x)^n,~~~~~~~x\in[0,1].$$ It is not difficult to check that
\begin{equation}\label{e:3.2}
h(x)=\int_0^1g_\alpha(x)w_n(\alpha)d\alpha,~~~~~~~x\in[0,1],
\end{equation}
where $g_\alpha(x)$ is as defined in (\ref{e:3.1}). By (\ref{e:2.1}), for any $X\in\LL^2$ we have
\begin{equation}\label{e:3.3}
\maxvar_n(X)=\int_{-\infty}^0[h(\PP_0(X>t))-1]dt+\int_0^{+\infty}h(\PP_0(X>t))dt.
\end{equation}
So by (\ref{e:3.2}), (\ref{e:3.3}) and Lemma \ref{l:Choquet}, together with Fubini's theorem and the fact that $\int_0^1w_n(\alpha)d\alpha=1$, we get
\begin{align*}
\maxvar_n(X)&=\int_{-\infty}^0\int_0^1[f_\alpha(X>t)-1]w_n(\alpha)d\alpha dt+\int_0^{+\infty}\int_0^1f_\alpha(X>t)w_n(\alpha)d\alpha dt\\
&=\int_0^1\left[\int_{-\infty}^0[f_\alpha(X>t)-1]dt+\int_0^{+\infty}f_\alpha(X>t)dt\right]w_n(\alpha)d\alpha\\
&=\int_0^1\cvar_\alpha(X)w_n(\alpha)d\alpha
\end{align*}
for any $X\in\LL^2$, as desired.\\

\noindent\textbf{Remark.}~Theorem \ref{t:mixedCVaR} says that $\maxvar_n(\cdot)$ is a continuous convex combination of the CVaR measure, its coherency in basic sense follows from Proposition 2.1 of Ang et al \cite{Ang-Sun-Yao2017}, and its averseness follows from the averseness of the CVaR~(Proposition 4.4 of Ang et al \cite{Ang-Sun-Yao2017}) together with the basic property of integral. Therefore, Theorem \ref{t:mixedCVaR} can actually provide an alternative proof of the coherency and averseness of $\maxvar_n(\cdot)$.

\section{The risk envelope of MAXVAR}
Since $$\maxvar_n(\cdot)=\int_0^1\cvar_\alpha(\cdot)w_n(\alpha)d\alpha$$ is a coherent risk measure on $\LL^2$, by the dual representation theorem (Rockafellar \cite{Rockafellar2007}), there exists a unique, nonempty, convex and closed set $\QQ_n\subseteq\LL^2$, called ``the risk envelope of $\maxvar_n(\cdot)$'' such that
$$\maxvar_n(X)=\sup\limits_{Q\in\QQ_n}\EE(XQ)$$ for any $X\in\LL^2$.

In this section we aim at characterizing the risk envelope of $\maxvar_n(\cdot)$''. First recall the following well-known result for the discrete convex combination of the  CVaR measure, which can be found in Rockafellar \cite{Rockafellar2007} and whose proof can be found in Ang et al \cite{Ang-Sun-Yao2017}.
\begin{prop}\label{p:mix_discrete}
Let $\RRR(\cdot)=\sum\limits_{i=1}^n\lambda_i\cvar_{\alpha_i}(\cdot)$ with positive weights $\lambda_i$ adding up to $1$. Then $\RRR$ is a coherent risk measure in the basic sense and its risk envelope is
$$\left\{\sum\limits_{i=1}^n\lambda_iQ_i:~0\leq Q_i\leq\frac{1}{1-\alpha_i},~\EE(Q_i)=1,~\forall i=1,2,\cdots,n\right\}.$$\\
\end{prop}

A continuous version of Proposition \ref{p:mix_discrete} gives the risk envelope of MAXVAR as follows.
\begin{thm}\label{51} The risk envelope of MAXVAR is
\begin{equation}\label{e:4.1}
\QQ_n:=\cl\left\{\int_0^1Q_\alpha w_n(\alpha)d\alpha,~0\leq Q_\alpha\leq\frac{1}{1-\alpha},~\EE(Q_\alpha)=1,~\forall\alpha\in[0,1)\right\},
\end{equation}
where $$w_n(\alpha):=n(n-1)(1-\alpha)\alpha^{n-2}~~~~~~\alpha\in[0,1]$$ is the weight function~($0^0$ is defined to be 1), and ``$\cl$'' stands for the closure in $\LL^2$.
\end{thm}

\pf
 Note that the integration ``$\int_0^1Q_\alpha w_n(\alpha)d\alpha$'' in (\ref{e:4.1}) is defined pointwise. That is, $Y=\int_0^1Q_\alpha w_n(\alpha)d\alpha$ means $Y(\omega)=\int_0^1Q_\alpha(\omega)w_n(\alpha)d\alpha$ for any $\omega\in\Omega$. Since $0\leq Q_\alpha\leq\dfrac{1}{1-\alpha}$ for any $\alpha\in[0,1)$, we have
$$0\leq\int_0^1Q_\alpha(\omega)w_n(\alpha)d\alpha\leq\int_0^1n(n-1)\alpha^{n-2}d\alpha=n$$ for any $\omega\in\Omega$. Therefore, $\QQ_n\subseteq\LL^\infty\subseteq\LL^2$. In addition, we can check that
\begin{align}
\maxvar_n(X)&=\int_0^1\cvar_\alpha(X)w_n(\alpha)d\alpha\nonumber\\
&=\sup\left\{\EE\left(X\int_0^1Q_\alpha w_n(\alpha)d\alpha\right):~\int_0^1Q_\alpha w_n(\alpha)d\alpha\in\QQ_n\right\}\label{52}
\end{align}
for any $X\in\LL^2$. Furthermore, it is easy to check the convexity of $\QQ_n$. Since $\QQ_n$ is closed in $\LL^2$, it follows from the dual representation theorem that Formula (\ref{52}) implies that (\ref{e:4.1}) is the risk envelope of $\maxvar_n(\cdot)$.
\qed\\

\noindent{\bf Acknowledgment.} We would like to thank the anonymous referees for their useful suggestions,
which are of great help for improving the manuscript. The research of Jie Sun is partially supported by  Australian Research Council under Grant DP160102819. The research of Qiang Yao is partially supported by grants from National Science Foundation of China (No.11201150 and No.11126236) and the 111 Project~(No.B14019).


\begin{thebibliography}{10}
\providecommand{\url}[1]{{#1}}
\providecommand{\urlprefix}{URL }
\expandafter\ifx\csname urlstyle\endcsname\relax
  \providecommand{\doi}[1]{DOI~\discretionary{}{}{}#1}\else
  \providecommand{\doi}{DOI~\discretionary{}{}{}\begingroup
  \urlstyle{rm}\Url}\fi


\bibitem{Ang-Sun-Yao2017} M. Ang, J. Sun and Q. Yao (2017) On the dual representation of coherent risk measures, {\em Ann. Oper. Res.} to appear. DOI: 10.1007/s10479-017-2441-3.

\bibitem{Cherny-Madan2009} A. Cherny and D. Madan (2009) New measures for performance evaluation, {\em Rev. Finan. Stud.} \textbf{22}, 2571-2606.

\bibitem{Cherny-Orlov2009} A. Cherny and D. Orlov (2011) On two approaches to coherent risk contribution, {\em Math. Finance} \textbf{23}, 557-571.

\bibitem{Follmer-Schied2002}H. F\"ollmer and  A. Schied (2004).{\em Stochastic Finance~(2nd Edition).} Walter de Gruyter, Berlin, Germany.

\bibitem{Rockafellar2007} R. T. Rockafellar (2007) Coherent approaches to risk in optimization under uncertainty, {\em Tutorials in Operations Research, INFORMS}, 38-61.

\bibitem{Rockafellar-Uryasev2000}  R.T. Rockafellar and  S. Uryasev (2000). Optimization of conditional value-at-risk. {\em Journal of Risk}, {\bf 2(3)}, 21-42.


\end{thebibliography}
\end{document}